\documentclass[twocolumn,prc,showpacs,preprintnumbers,superscriptaddress]{revtex4-1}

\usepackage{graphicx}
\usepackage{amsmath} 
\usepackage{amssymb}
\usepackage{amsfonts}
\usepackage{mathrsfs}
\usepackage{units}
\usepackage{multirow}

\usepackage{dcolumn}
\usepackage{bm}
\usepackage{rotating} 

\newcommand{\gras}[1]{\boldsymbol{#1}}

\begin{document}

\title{Alpha Decay in the Complex Energy Shell Model}

\author{R. Id Betan}

\affiliation{Department of Physics and
  Astronomy, University of Tennessee, Knoxville, Tennessee 37996, USA}
\affiliation{Physics Division, Oak Ridge National Laboratory, P.O. Box
  2008, Oak Ridge, Tennessee 37831, USA}
\affiliation{Departamento de Qu\'{\i}mica y F\'{\i}sica, FCEIA(UNR) - Instituto de F\'{\i}sica Rosario (CONICET), Av. Pellegrini 250, 2000 Rosario, Argentina}

\author{W. Nazarewicz}

\affiliation{Department of Physics and
  Astronomy, University of Tennessee, Knoxville, Tennessee 37996, USA}
\affiliation{Physics Division, Oak Ridge National Laboratory, P.O. Box
  2008, Oak Ridge, Tennessee 37831, USA}
\affiliation{Institute of
  Theoretical Physics, University of Warsaw, ul. Ho\.za 69, 00-681
  Warsaw, Poland}

\date{\today}

\begin{abstract} 
\begin{description} 
\item[Background] Alpha emission from a nucleus is a fundamental decay process in which the alpha particle formed inside the nucleus tunnels out through the potential barrier.
\item[Purpose] We describe alpha decay of $^{212}$Po and $^{104}$Te by means of the configuration interaction approach.
\item[Method] To compute the preformation factor and penetrability, we use the complex-energy shell model with a separable $T$=1 interaction. The single-particle space is expanded in a Woods-Saxon basis that consists of bound and  unbound resonant states. Special attention is paid to the treatment of the norm kernel appearing in the definition of the formation amplitude that guarantees the normalization of the channel function. 
\item[Results] Without 
explicitly considering the alpha-cluster component in the
wave function of the parent nucleus, we reproduce the experimental alpha-decay width of $^{212}$Po and predict an upper limit  of $T_{1/2}=5.5 \times 10^{-7}$\,sec for the half-life of $^{104}$Te.
\item[Conclusions] The complex-energy shell model in a large valence configuration space is capable of providing a microscopic description of the alpha decay of heavy nuclei having two valence protons and two valence neutrons outside the doubly magic core. The inclusion of proton-neutron interaction between the valence nucleons is likely to shorten the predicted half-live of $^{104}$Te.
\end{description}
\end{abstract}


\maketitle

\section{Introduction} \label{sec.introduction}

According to Gamow theory of alpha decay \cite{1928Gamow,1928Condon}, this fundamental radioactive decay   can be considered as a two-step process \cite{1998Lovas,2003Hodgson,2010Delion}. In the first step,  an alpha  cluster is formed inside the parent nucleus.   
The resulting alpha particle  resides  in a metastable state of an average potential  of the daughter system.  In the second step, the particle tunnels  through the potential barrier. Each step requires different theoretical treatment. To compute the preformation factor that  describes the alpha formation probability, one needs to evaluate the overlap integral involving wave functions of the parent and daughter nuclei, and that of the alpha particle. The estimate of the penetration probability requires a careful treatment of 
the  resonance state.

The commonly used  formulation of the alpha-decay problem employs the R-matrix expression \cite{1954Thomas,1957Mang}   
\begin{equation}\label{Grmatrix}
\Gamma_L = 2 P_L \gamma_L^2
\end{equation}
for  the absolute width. In this formalism, the first stage (formation of alpha particle with angular momentum $L$) is given by the reduced width $\gamma_L^2$,  while the second stage (decay) is expressed by means of the penetrability $P_L$. 
Alternatively, the absolute width can be obtained from  the general reaction-theory expression \cite{1972Arima,1974Arima,1976Fliessbach-jpg,1977Fliessbach}
\begin{equation}\label{Goverlap} 
\Gamma_L = S_L \Gamma_L^\text{sp},
\end{equation}
where $S_L$ is the alpha-spectroscopic factor and  $\Gamma_L^\text{sp}$ is the
single-particle (s.p.)  decay width.

Historically, expression (\ref{Grmatrix}) was derived in 1954 by Thomas \cite{1954Thomas} using  the time-independent R-matrix theory of nuclear reactions. 
In 1957, Mang \cite{1957Mang} developed the alpha-decay formalism based on the time-dependent perturbation theory. He made the connection with the shell model  and succeeded in expressing the alpha-decay formation amplitude in a basis of s.p. states. As shown in Refs.~\cite{1963Zeh,1964Mang}  formulations of Thomas and Mang are formally equivalent; there are, however, many differences when it comes to practical implementations.

The reduced width
calculated in a  shell-model configuration expressed in the harmonic oscillator (h.o.) basis
is too  small. This can be partly cured by  means of configuration mixing involving extended shell-model spaces \cite{1961Harada,1979Tonozuka} as each admixed configuration contributes coherently to $\gamma_L^2$.  To improve asymptotic properties of s.p. wave functions, the particle continuum was taken into account \cite{1983Janouch} by considering h.o. expansion \cite{1988Dodig-Crnkovic} or within a Woods-Saxon (WS) basis consisting of bound and outgoing single-particle resonant (Gamow) states \cite{1993Lenzi,2000Delion}.  The configuration mixing calculations of  Refs.~\cite{1979Tonozuka,1993Lenzi} in the valence space of $^{212}$Po assumed the seniority-zero (pairing vibrational) wave functions 
obtained by considering the monopole pairing interaction between like nucleons.
However, all these improvements were not sufficient to reproduce the experimental alpha decay in $^{212}$Po. It is only after the valence proton-neutron interaction had been considered together with  a generalized wave function expressed as a combination of cluster and shell model components \cite{1989Okabe} that theoretical  and  experimental widths could be reconciled \cite{1992Varga}.

The R-matrix expression for the width (\ref{Grmatrix}) depends on the channel radius $R$. This radius should be chosen large enough so that the alpha-daughter interaction  in the external region   is given by the   Coulomb force alone \cite{2010Descouvemont}. The infinite range of the Coulomb force implies, however,  that the asymptotic behavior of the R-matrix expression is  reached only at  large values of $R$, at which  the  asymptotic behavior of the shell-model s.p. basis (h.o. basis in most applications) used to calculate $\gamma_L^2(R)$,  does matter. Due to the mismatch between the internal part of the s.p. wave function (well described in the h.o. basis) and the asymptotic part (poorly or not described in the h.o. basis), rather small changes in $R$ may  produce appreciable variations in penetrability. Physically, the reason for this sensitivity is the fact that the alpha cluster is formed in the surface region  of the nucleus in which the coupling to the  alpha continuum that impacts the radial behavior of the formation amplitude  is important \cite{1983Janouch}. Consequently, the absolute R-matrix width depends in general on the channel radius \cite{1991Insolia,2010Delion}, and this is an obvious drawback of the method \cite{2010Descouvemont}. 

Our renewed interest in the alpha-decay problem is stimulated by the recent experimental data above the doubly-magic $^{100}$Sn \cite{2005Janas,2006Liddick} that demonstrate the presence of  very fast alpha decays. Indeed, the observed enhancement of the reduced widths of $^{105,106}$Te relative to 
$^{213,212}$Po is two-to-three, thus confirming earlier expectations \cite{1965Macfarlane} of ``superallowed" alpha decays in this region due to the large  overlaps of valence s.p. shell model proton and neutron wave functions. 
Our  long-term goal  is to estimate  alpha preformation factors in nuclei above $^{208}$Pb  and  $^{100}$Sn  by using  large valence s.p. spaces, including positive-energy Gamow states of a finite-depth WS potential \cite{1993Lenzi,2000Delion}. In this study, we focus on  
$^{212}$Po and $^{104}$Te  nuclei having two valence protons and two valence neutrons outside  doubly-magic cores.

Our paper is organized as follows. Section~\ref{sec.formalism} briefly describes the alpha-decay formalism used in this work, with  special emphasis on approximations used to describe wave functions of parent and daughter nuclei. Section \ref{sec.model} deals with the approximations employed and parameters used. In particular, we discuss the sensitivity of the calculated spectroscopic factor to the parameters defining the shifted Gaussian basis that is used to compute the normalization of the channel function. In Sec.~\ref{sec.applications.ho} we study the sensitivity of the reduced alpha width in $^{212}$Po  on the choice of s.p. basis used. In Section \ref{sec.aw} we discuss the absolute alpha-decay width  of $^{212}$Po and in Sec.~\ref{sec.applications.compare} we compare it with the absolute width of the superallowed alpha emitter  $^{104}$Te. Finally, the main conclusions of this work are summarized in Sec.~\ref{sec.conclusions}.

\section{Formalism}\label{sec.formalism}
In this section, we discuss the  R-matrix (\ref{Grmatrix}) and 
spectroscopic factor (\ref{Goverlap})
expressions for the decay width. 
The connection between the two formulations is given in Ref.~\cite{1972Arima}. We also discuss the so-called delta-approximation for the formation amplitude.

\subsection{R-matrix expression for the decay width}\label{subsec.formalism.r}

Within the R-matrix theory \cite{1954Thomas,1957Mang,1963Zeh},  the absolute width is given by Eq.~(\ref{Grmatrix})
with  $P_L(R)$ being the barrier penetrability  and $\gamma_L(R)$  -- the reduced width amplitude  \cite{1960Lane}. While both quantities strongly depend on the value of the channel radius  $R$, the absolute width should be $R$-independent. 

For $P_L(R)$ we  use the standard expression \cite{1954Thomas}:
\begin{equation} 
 P_L(R) = \frac{kR}{|H_L^+(\eta, k R)|^2}, 
\end{equation}
where $k$ is given by the  alpha energy 
 $E_\alpha=\frac{ \hbar^2 \; k^2}{2 \; \mu}$, obtained
from the experimental $Q_\alpha$ value 
by correcting for  electron screening;
$\mu=\frac{m_d m_\alpha}{m_d + m_\alpha}$ is the reduced mass of alpha particle
with $m_d$ being the mass of the daughter nucleus;
$H_L^+(\eta, kR)$ is the outgoing spherical Coulomb-Hankel function; and  $\eta=\frac{2 Z_d \mu e^2 }{\hbar^2 k}$ is the Sommerfeld Coulomb parameter.

The reduced width amplitude  $\gamma_L(R)$ may be written in terms of the formation amplitude $g_L(R)$ \cite{1961Harada,2010Delion}: 
\begin{equation} \label{eq.gamma}
 \gamma_L = \sqrt{\frac{\hbar^2 R}{2 \mu}} g_L(R),
\end{equation}
with 
\begin{eqnarray} \label{eq.flfirst}
 g_L(R)&=& \int d\Omega_R \int d\xi_\alpha \int d\xi_D \; \nonumber \\
	 && \Phi^P_{J M}\;  
 	 \mathcal{A} \left[  \phi_\alpha(\xi_\alpha) \; \Psi^D_j(\xi_D)\; Y_L(\hat{R}) \right]^*_{J M},
\end{eqnarray}
where $\phi_\alpha$ is the normalized wave function of the alpha particle with zero angular momentum, $Y_{L M_L}$ is the angular part of the center-of-mass  (c.o.m.) motion of the alpha particle, $\Psi^D_{j m_j}$ is the wave function of the daughter nucleus, and $\Phi^P_{J M}$ is the wave function of the parent nucleus.
The coordinates $\xi_\alpha$ and $\xi_D$ are the intrinsic coordinates of the  alpha particle and  daughter nucleus, respectively. 
All wave functions are normalized in terms of the internal and c.o.m. coordinates \cite{1960Lane}.
By construction, the parent and daughter wave functions   are antisymmetric. 
The antisymmetrization with respect to inter-fragment nucleons
is done by means of the  operator $\mathcal{A}$. Its action can be approximated by means of a factor $\left[ \binom{N_v}{2}\binom{Z_v}{2} \right]^{1/2}$ \cite{1960Lane,1960Mang,1961Harada}, with $N_v$ and $Z_v$ being, respectively,  the numbers of valence neutrons and protons in the parent nucleus.

For the internal alpha-particle wave function we take the standard Gaussian ansatz \cite{1990Bayman,1993Lenzi}:
\begin{eqnarray}
 \phi_\alpha(\rho_1 \rho_2 \rho_3, \sigma_1 \sigma_2 \sigma_3 \sigma_4) &=&
       \phi(\rho_1 \rho_2 \rho_3) \chi_{00}(\sigma_1 \sigma_2) \chi_{00}(\sigma_3 \sigma_4),  \nonumber \\
  \chi_{00}(\sigma_1 \sigma_2) &=& [\chi_{1/2}(\sigma_1) \chi_{1/2}(\sigma_2)]_{00},  \nonumber \\
  \phi(\rho_1 \rho_2 \rho_3) &=& \left( \frac{8 \beta}{\pi} \right)^{9/4} e^{-4 \beta (\rho_1^2 + \rho_2^2 + \rho_3^2)}.
\end{eqnarray}
The parameter $\beta=\frac{9}{64 r_\alpha^2}$=0.057\,fm$^{-2}$ depends on  the  root-mean-square alpha radius $r_\alpha$=1.57\,fm \cite{1990Bayman}.

The transformation between the intrinsic $\xi_\alpha=\{ \bm{\rho}_1, \bm{\rho}_2, \bm{\rho}_3 \}$ and nucleonic  $\{\gras{r_i}\}$ ($i$=1,2,3,4) coordinates reads:
\begin{eqnarray} \label{eq.rel}
 \gras{\rho}_1&=& \frac{\gras{r}_1 - \gras{r}_2}{\sqrt{2}}, \nonumber \\
 \gras{\rho}_2&=& \frac{\gras{r}_3 - \gras{r}_4}{\sqrt{2}}, \\
 \gras{\rho}_3&=& \frac{(\gras{r}_1 + \gras{r}_2) - (\gras{r}_3 + \gras{r}_4)}{2}\nonumber,
\end{eqnarray}
and
\begin{equation}\label{com}
\gras{R}= \frac{\gras{r}_1+\gras{r}_2+\gras{r}_3+\gras{r}_4}{4}
\end{equation}
is the c.o.m. coordinate of alpha particle. 
Let us denote the spherical components of intrinsic coordinates by $\gras{\rho}_i=(\rho_i,\tilde{\theta}_i,\tilde{\varphi}_i)$. Assuming $\theta_R=\varphi_R=0$, the nucleonic coordinates  can be written as:
\begin{eqnarray}\label{eq.rhoandr}
 4 r^2_{1,2} &=& 4R^2 + \rho^2_3 + 2 \rho^2_1 \pm 2\sqrt{2}  \rho_3 \rho_1  \cos \tilde{\theta}_{31} \nonumber \\
        &&      + 4  R \left( \rho_3 \cos \tilde{\theta}_3 \pm \sqrt{2} \rho_1
        \cos \tilde{\theta}_1 \right),   \nonumber \\
4 r^2_{3,4} &=& 4R^2 + \rho^2_3 + 2 \rho^2_2 \mp 2\sqrt{2}\rho_3\rho_2 \cos \tilde{\theta}_{32} \nonumber \\
         &&     - 4 R \left( \rho_3 \cos \tilde{\theta}_3 \pm \sqrt{2} \rho_2 \cos \tilde{\theta}_2 \right),   
\end{eqnarray}
where  $\tilde{\theta}_{ij} = \tilde{\theta}_j - \tilde{\theta}_i$, and
\begin{eqnarray}\label{eq.cos}
  \cos \theta_{1,2} &=& \frac{2R  + \rho_3 \cos \tilde{\theta}_3 \pm \sqrt{2} \rho_1 \cos \tilde{\theta}_1}{2r_{1,2}}, \nonumber \\
  \cos \theta_{3,4} &=& \frac{2R  - \rho_3 \cos \tilde{\theta}_3 \pm \sqrt{2} \rho_2 \cos \tilde{\theta}_2}{2r_{3,4}}. 
\end{eqnarray}

This paper deals with  g.s.$\rightarrow$g.s. alpha decays to the magic  daughter nucleus. Assuming the  seniority-zero wave function, the corresponding formation amplitude is \cite{1961Harada,1964Mang}
\begin{equation}\label{eq.f0}
 F_0(R) =  \frac{\sqrt{8}}{16 \pi^{3/2}}  \sum_{\nu_n,\nu_p}  (-)^{l_n+l_p} b_{\nu_n,\nu_p} \hat{j}_n  \hat{j}_p I_{v_n,\nu_p}(R),
\end{equation}
where
\begin{eqnarray}\label{eq.inp}
    I_{\nu_n,\nu_p}(R) &=&   \int d\gras{\rho}_1 d\gras{\rho}_2 d\gras{\rho}_3 
 	      \phi(\rho_1 \rho_2 \rho_3)   \nonumber \\
    &\times&    \frac{u_{\nu_n}(r_1)}{r_1} \frac{u_{\nu_n}(r_2)}{r_2} P_{l_n}(\cos\theta_{12})  \\
    &\times&    \frac{u_{\nu_p}(r_3)}{r_3}  \frac{u_{\nu_p}(r_4)}{r_4} 
    P_{l_p}(\cos\theta_{34}), \nonumber
\end{eqnarray}
with $\theta_{ij}=\theta_j - \theta_i$, $\nu= \{n,l,j\}$, and $u_{\nu}(r)$ being s.p.  radial wave functions.
The factor $\sqrt{8}$ comes from the Jacobian of the transformation between  
the nucleonic coordinates $\{\gras{r_i}\}$  and the 
internal and  c.o.m.   coordinates \cite{1965Eichler,2010Delion}. 
In Eq.~(\ref{eq.inp}) and in the following, the s.p. indices $1,2$ refer to neutrons while $3,4$ refer to protons. 
The coefficients $b_{\nu_n,\nu_p}$ are the shell-model four-particle wave function amplitudes.

\subsection{Delta-function approximation}\label{subsec.formalism.delta}

In the calculation of alpha-decay rates based on h.o. wave functions, it was noticed \cite{1960Mang} that the relative rates   change little with the oscillator length $b_\text{h.o.}$ of the basis. Using this argument,  Mang proposed to take  $\beta \gg 1/b^2_\text{h.o.}$. In this limit, the expression for the formation amplitude  can be simplified (see also  Ref.~\cite{1958Brussaard}).
In the literature, this is known as delta-function approximation
\cite{1963Rasmussen}.

In practice, one  assumes that the alpha particle wave function 
is  constant inside a small volume of radius $s_\alpha =2.34$\,fm \cite{1963Rasmussen} and zero outside. 
Within this approximation  $\gras{\rho}_i=0$; hence, it immediately follows from Eqs.~(\ref{eq.rhoandr})  that $\gras{r}_1=\gras{r}_2=\gras{r}_3=\gras{r}_4=\gras{R}$ \cite{1957Devons,1963Rasmussen},  and the formation amplitude reduces to
\begin{equation}
 F^\delta_0(R) = \frac{\sqrt{8}}{16 \pi^{3/2}}  \left( \frac{4\pi s^ 3_\alpha}{3} \right)^{3/2}  
	\left( \sum_{\nu_n} I^n_{\nu_n} \right) \left( \sum_{\nu_p} I^p_{\nu_p} \right),
\end{equation}
with
\begin{equation}\label{Bcorr2}
   I^\tau_\nu = (-)^{l_\nu} b^\tau_\nu  \hat{j}_\nu   B_\nu \frac{u^2_{\nu\tau}(R)}{R^2}, 
\end{equation}
where $\tau=n, p$. The correction factor  $B_\nu$ depends on the relative angular momentum \cite{1963Rasmussen}:
\begin{equation}\label{Bcorr}
B_\nu=1-0.013 l_\nu(l_\nu+1).
\end{equation}

\subsection{Four-particle amplitudes} \label{sec.fpa}
For the g.s. alpha decay of $^{212}$Po and $^{104}$Te, we are going 
to assume that the four valence nucleons move around the rigid, doubly-magic core. 
The parent-nucleus  wave function is approximated by a product of two-neutron and two-proton  seniority-zero  states:
\begin{equation}
  |\Phi^P_{J=0,M=0}\rangle= |\Psi_{2n,{00}} \rangle \otimes |\Psi_{2p,{00}} \rangle,
\end{equation}
where
\begin{equation}\label{twop}
 |\Psi_{2\tau,0} \rangle = \sum_{\nu} X_{\nu}^{\tau} |\nu \nu,00 \rangle ,
\end{equation}
$|\nu \nu,00 \rangle=\frac{ [a_{\nu}^\dagger a_{\bar\nu}^\dagger]_{00} }{ \sqrt{2} } |0_\tau \rangle$, and $|0\rangle=|0_n \rangle\otimes |0_p \rangle$ is the shell-model vacuum representing the  $^{208}$Pb or $^{100}$Sn g.s. wave function. The four-particle amplitudes $b_{\nu_n,\nu_p}$ in (\ref{eq.f0})  can thus be written in a separable form:
\begin{equation}
 b_{\nu_n,\nu_p} = X_{\nu_n}^n X_{\nu_p}^p.
\end{equation}

\subsection{Alpha decay spectroscopic factor}\label{subsec.formalism.s}
Based on the general theoretical arguments \cite{1972Arima,1974Arima,1976Fliessbach-jpg,1977Fliessbach}, the absolute width can be expressed as a product of the alpha-particle spectroscopic factor and the single particle width, see Eq.~(\ref{Goverlap}). The spectroscopic factor  $S_L$  contains information about the probability of forming an alpha cluster in the parent system. Since the alpha particle, when formed,  occupies the resonant state, the s.p. width can be obtained from the so-called current expression \cite{1961Humblet,2000Barmore,2010Delion}:
\begin{equation} \label{eq.widthsp1}
 \Gamma_L^\text{sp} = i\frac{\hbar^2 }{2 \mu} \frac{u'^*_L(R)\; u_L(R) - u'_L(R)\; u^*_L(R)}{\int \; |u_L(R)|^2 \; dR}, 
\end{equation}
where the Gamow function $u_L(R)$ is obtained as a solution of the Schr{\"o}dinger equation with  outgoing boundary condition.
When the imaginary part of the complex energy eigenvalue $\mathcal{E}_\alpha=\frac{ \hbar^2 \; k^2}{2 \; \mu}$ is small, which is always the case for the considered g.s. alpha emitters, one can approximate (\ref{eq.widthsp1}) with \cite{2004Kruppa}:
\begin{equation} \label{eq.gammasp}
 \Gamma_L^{sp} = \frac{\hbar^2 \Re(k)}{\mu} \frac{|u_L(R)|^2}{|H^{+}_L(\eta,k R)|^2 }. 
\end{equation}
The s.p. width obtained in this way should be identical to the value
$-2$Im($\mathcal{E}_\alpha$) given by the imaginary part of the Gamow resonance energy, if the latter is computed with a sufficient precision.

The conventional alpha spectroscopic factor as introduced in Ref.~\cite{1972Arima} is defined by 
\begin{equation} \label{eq.sbaket}
  S_L= | \langle \mathcal{A} \left[ \phi_\alpha(\xi_\alpha) \; \Psi^D_j(\xi_D)\; \psi_L(\gras{R}) \right]_{J M} | \Phi^P_{J M} \rangle |^2,
\end{equation}
where $\psi_{LM}(\gras{R})= \frac{u_L(R)}{R} Y_{LM}(\hat{R})$ represents  the relative motion alpha particle with respect to the daughter.
In terms of the formation amplitude, $S_L$ reads \cite{1998Lovas,1976FliessbachMang,2010Delion}:
\begin{equation} \label{eq.sint}
 S_L = \int_0^\infty g_L^2(R) R^2  dR.  
\end{equation}

\subsection{Modified spectroscopic factor} \label{app.norm}

Since the formation amplitude Eq. (\ref{eq.flfirst}) represents the overlap of the parent wave function with the daughter-alpha product state,  one would be tempted to associate it with the probability amplitude that in the parent wave function $\Phi^P_{JM}$ an alpha particle $\phi_\alpha$ and a daughter nucleus $\Psi^D_{j m_j}$ are  at a distance $R$. The value of $S_L$  would then be associated with the total probability of formation of an alpha particle. However, the fundamental problem with this interpretation is that the channel function 
$\mathcal{A} \left[ \phi_\alpha(\xi_\alpha) \; \Psi^D_j(\xi_D)\; \psi_L(\gras{R}) \right]_{J M}$ is not properly normalized \cite{1975Fliessbach,1976Fliessbach-jpg,1977Fliessbach,1978Fliessbach,1980Watt,1987Blendowske,1998Lovas}. 

The properly defined spectroscopic factor (sometimes referred to as ``the amount of clustering") \cite{1975Fliessbach,1976Fliessbach,1976Fliessbach1,1987Beck,1992VargaLovas,1992Varga} is given by
\begin{equation}\label{eq.s2}
 \mathcal{S}_L= \int_0^\infty G_L^2(R) R^2  dR, 
\end{equation}
where
\begin{equation}\label{eq.gbig}
 G_L(R)= \int  \mathcal{N}_L^{-1/2}(R,R') \; g_L(R')R'^2  \;dR'
\end{equation}
is the modified formation amplitude. The norm kernel $\mathcal{N}_L$ appearing in Eq.~(\ref{eq.gbig}) is \cite{1976Fliessbach1}
\begin{eqnarray} \label{eq.nk}
&& \mathcal{N}_L(R,R') =  \\
&&     \langle \mathcal{A} 
                           \frac{\delta(R_\alpha - R)}{R^2} \phi_\alpha 
                           \left[ Y_L \Psi_j^D \right]_J 
                          |
             \mathcal{A} 
                           \frac{\delta(R_\alpha - R')}{R'^2} \phi_\alpha 
                           \left[ Y_L \Psi_j^D \right]_J  
       \rangle. \nonumber  
\end{eqnarray}
The presence of the norm kernel $\mathcal{N}$ effectively enhances the spectroscopic factor by one-to-two orders of magnitude \cite{1979Conze,1980Watt,1987Blendowske,1992VargaLovas,1992Varga}. 

To compute  $\mathcal{N}_L^{-1/2}(R,R')$, we expand the eigenfunctions of the norm kernel in an orthonormalized shifted Gaussian basis (SGB) \cite{1976Fliessbach1},
\begin{equation} \label{eq.tildeF}
 \tilde{F}_L(R,R_k) = \sum_{k'}  \left( N^{-1/2}_F \right)_{k k'} \; F_L(R,R_{k'}), 
\end{equation}
with $R_k$ equidistant mesh points in the interval $(0,R_{\textrm{max}})$ and $k=1,\dots,M$, where $M$ is the dimension of the basis. The SGB is given by
\begin{equation}\label{eq.sgb}
 F_L(R,R_k) = 4 \pi \left( \frac{8 \beta'}{\pi} \right)^{3/4} e^{-4 \beta' (R^2 + R_k^2)}  i^L j_L(-i8 \beta' R R_k),
\end{equation}
while the SGB overlap $(N_F)_{k k'}$ is given by 
\begin{eqnarray} \label{eq.nf}
 (N_F)_{k k'} &=& \int  F^*_L(R,R_k)  F_L(R,R_{k'}) R^2 dR  \\
            &=& 4 \pi e^{-2 \beta' (R_k^2 + R^2_{k'})}   i^L j_L(-i 4 \beta' R_k R_{k'}). \nonumber
\end{eqnarray}
Using the SGB  overlaps, the eigenvalue equation for the norm matrix can be expressed in the form:
\begin{equation}\label{fnc}
 \sum_{k'}^M \mathcal{N}^{\tilde{F}}_{kk'} \; c^\nu_{k'} = n_\nu \;  c^\nu_k,
\end{equation}
where
\begin{equation}\label{NtF}
 \mathcal{N}^{\tilde{F}}_{kk'} = \sum_{nn'}^M \; \left( N^{-1/2}_F \right)_{k n} \; \mathcal{N}^F_{nn'} \; \left( N^{-1/2}_F \right)_{n' k'}
\end{equation}

For $\beta'=4 \beta$, the core-projected norm $\mathcal{N}^F$ in Eq.~(\ref{NtF}) reduces to a simple expression \cite{1992Varga,1976Fliessbach1,1966Brink}:
\begin{equation} \label{eq.nkkp}
  \mathcal{N}^F_{kk'} =  \left( \langle \psi^{(\nu),L}_k | \psi^{(\nu),L}_{k'} \rangle \right)^2 \;
                         \left( \langle \psi^{(\pi),L}_k | \psi^{(\pi),L}_{k'} \rangle \right)^2 
\end{equation}
where
\begin{eqnarray}\label{eq.nkkp2}
\langle \psi^{(\mu),L}_k | \psi^{(\mu),L}_{k'} \rangle & = &  
                \langle \phi^L_k | \phi^L_{k'} \rangle \nonumber \\
&-& \sum_{nlj\tau \in {\rm core}} \delta_{lL} \langle \phi^l_k | R_{nlj} \rangle \langle R_{nlj} | \phi^l_{k'} \rangle  
\end{eqnarray}
with $\phi^L_k(R) =  F_L(R,R_k) (\beta'\rightarrow \beta)$ 
and $R_{nlj}(R)=u_{nlj}/R$ are the radial s.p. wave functions of the core.

In terms of eigenstates $c^\nu_k$ of (\ref{fnc}), the spectral representation of the norm kernel can be written as:
\begin{equation}
 \mathcal{N}_L^{-1/2}(R,R') = \sum_{\nu \atop (n_\nu > n_{\rm min})}  \; n_\nu^{-1/2} \; u^{L*}_\nu(R) \; u^L_\nu(R'),
\end{equation}
where the  eigenfunctions $u^L_\nu(R)$ of the norm kernel are
\begin{equation} \label{eq.ur}
 u^L_\nu(R) = \sum_k^M \; c^\nu_k \; \tilde{F}_L(R,R_k),
\end{equation}
and  $n_{\rm min}$ represents the usual cutoff on the eigenvalue of the norm kernel.  The final expression for the  modified formation amplitude in the normalized SGB becomes \cite{1976Fliessbach1}:
\begin{equation} \label{eq.gbig2}
 G_L(R) =\sum_{\nu \atop (n_\nu > n_{\rm min})} \; n_\nu^{-1/2} \; u^L_\nu(R) \; g^L_\nu   
\end{equation}
with
\begin{equation} \label{eq.gnu}
 g^L_\nu = \int  u^L_\nu(R) \; g_L(R) \;  R^2 dR.
\end{equation}

\section{The model} \label{sec.model}

\subsection{Single-particle space} \label{subsec.sp}

The s.p. space is spanned on resonant states of a WS+Coulomb average potential. The parameters of the s.p. Hamiltonian, namely
the WS potential depth $V_0$, spin-orbit potential depth   $V_\text{so}$, diffuseness $a$ ($=a_\text{so}$), radius $r_0$ ($=r_{0,\text{so}}$), and the radius of the uniform charge distribution $r_c$ defining the Coulomb potential  are listed in Table~\ref{table.ws}.
 \begin{table}[ht]
 \caption{\label{table.ws} Parameters of the average WS Hamiltonian used in this work to compute s.p. neutron and proton states  of  $^{208}$Pb and $^{100}$Sn cores.}
 \begin{ruledtabular}
 \begin{tabular}{ccccccc}
 Core   & $\tau$     & \mbox{$V_0$} & \mbox{$V_\text{so}$} & \mbox{$a$} & \mbox{$r_0$}  & \mbox{$r_c$} \\
     &                  & \mbox{(MeV)} &  \mbox{(MeV)}     &  \mbox{(fm)} &  \mbox{(fm)}   & \mbox{(fm)} \\
 \hline
\multirow{2}{*}{$^{208}$Pb}  & $n$   & 44.40          & 16.5                & 0.70      & 1.27      & \\
 & $p$  & 66.04           & 19.0                & 0.75      & 1.19        & 1.27              \\[2pt]
\multirow{2}{*}{$^{100}$Sn} & $n$   & 51.60          & 11.3                & 0.70      & 1.27       &\\
  & $p$  & 52.20          & 10.5                & 0.70      & 1.27          & 1.27 
 \end{tabular}
 \end{ruledtabular}
 \end{table}
The resulting neutron and proton s.p. energies for $^{208}$Pb and $^{100}$Sn are given in Tables~\ref{table.spPb} and \ref{table.spSn}, respectively. 
The nucleus $^{101}$Sb is proton-unbound; the values in Table~\ref{table.spSn} are generally consistent with systematics \cite{2002Isakov}. In particular, we predict
a very small splitting between the $0g_{7/2}$ and $1d_{5/2}$ neutron shells outside $^{100}$Sn,  and a  $0g_{7/2}$ g.s. in $^{101}$Sn as suggested by recent experiment \cite{2010Darby}.

 \begin{table}[htb]
 \caption{\label{table.spPb} The eigenstates (in MeV) of the s.p. Hamiltonian of Table~\ref{table.ws} for $^{208}$Pb calculated
with the Gamow solver {\sc anti} \cite{1995Ixaru}. The positive-energy eigenvalues represent Gamow resonances; their imaginary energies reflect nonzero particle width.}
 \begin{ruledtabular}
 \begin{tabular}{cc|cc}
   Orbit          & Neutrons & Orbit         & Protons \\
 \hline
  $1g_{9/2}$  & $-3.926$                  & $0h_{9/2}$  & $-3.784$ \\
  $0i_{11/2}$ & $-2.797$                   & $1f_{7/2}$  & $-3.542$  \\
  $2d_{5/2}$  & $-2.072$                   & $0i_{13/2}$ & $-1.844$ \\
  $0j_{15/2}$ & $-1.883$                   & $2p_{3/2}$  & $-0.690$ \\
  $3s_{1/2}$  & $-1.438$                   & $1f_{5/2}$  & $-0.518$ \\
  $2d_{3/2}$  & $-0.781$                   &  $2p_{1/2}$  &  $0.491-i0.200 \times 10^{-11}$ \\
  $1g_{7/2}$  & $-0.768$                   &  $1g_{9/2}$  &  $4.028-i 
  0.130 \times 10^{-7}$ \\
  $1h_{11/2}$ &  $2.251 - i0.026$ & $0i_{11/2}$ &  $5.434 - i0.992 \times 10^{-8}$  \\
  $0j_{13/2}$ &  $5.411 - i0.009$ &  $0j_{15/2}$ &  $5.960 - i0.115\times 10^{-7}$ \\
                    &                            & $2d_{5/2}$  &  $6.748 - i0.184\times 10^{-2}$  \\
                    &                            &$3s_{1/2}$  &  $7.843 - i0.367\times 10^{-1}$   \\
                    &                            & $1g_{7/2}$  &  $8.087 - i0.898\times 10^{-3}$  \\
                    &                            & $2d_{3/2}$  &  $8.530 - i0.284\times 10^{-1}$  \\
                    &                            &$1h_{11/2}$  &  $11.390- i0.215\times 10^{-1}$   \\
                    &                            &$0j_{13/2}$  &  $15.086 - i0.493\times 10^{-2}$   \\
                    &                            & $1h_{9/2}$  &  $15.964 - i0.393$ 
 \end{tabular}
 \end{ruledtabular}
 \end{table}

\begin{table}[htb]
\caption{\label{table.spSn} Similar as in Table~\ref{table.spPb} except for $^{100}$Sn.}
 \begin{ruledtabular}
\begin{tabular}{ccc}
  Orbit          & Neutrons &  Protons \\
\hline
 $0g_{7/2}$  & $-10.830$  & $2.669 - i0.207\times 10^{-7}$\\
 $1d_{5/2}$  & $-10.674$  & $2.869 - i0.963\times 10^{-5}$\\
 $2s_{1/2}$  & $-9.074$   &  $4.150 - i0595\times 10^{-2}$\\
 $1d_{3/2}$  & $-8.927$   &  $4.393 - i0.166\times 10^{-2}$\\
 $0h_{11/2}$ & $-5.793$  &  $7.280 - i0.110\times 10^{-2}$ \\
 $1f_{7/2}$  & $-2.346$                       & $9.649 -i 0.452$   \\
 $2p_{3/2}$  & $-1.531$                       & \\
 $2p_{1/2}$  & $-0.912$                       & \\
 $0h_{9/2}$  & $-0.641$                       & $12.012 -i 0.0736$ \\
 $1f_{5/2}$  & $-0.171$                       & \\
 $0i_{13/2}$ & $3.254 -i 0.132\times 10^{-2}$ & $15.572 -i 0.185$  
\end{tabular}
 \end{ruledtabular}
\end{table}

\subsection{Two-particle interaction} \label{subsec.force}

The correlated two particle wave functions $|\Psi_{2\tau,0} \rangle$ (\ref{twop}) have been obtained using a separable two-body  $T=1$ pairing interaction \cite{1971Bes}:
\begin{equation}\label{vres}
 \langle \nu \nu, 00 | V | \nu' \nu',00 \rangle = 
          -G_{\tau} f(\nu,\tau) f(\nu',\tau),
\end{equation}
where
\begin{equation}\label{fntau}
 f(\nu,\tau)= \frac{(-)^{l_\nu}}{\sqrt{2}} \langle j_\nu|| Y_0 || j_\nu \rangle I(\nu,\tau).
\end{equation}
In Eq.~(\ref{fntau}) we used the Condon-Shortley phase convention for $\langle j_\nu|| Y_0 || j_\nu \rangle$ and
\begin{equation}
 I(\nu,\tau)=\int u_{\nu \tau}^2(r) f_\tau (r) dr.
\end{equation}
For the radial form factor $f_\tau(r)$ we took  the derivative of the WS potential
multiplied by $r$:
\begin{equation}\label{ff}
 f_\tau (r)=\frac{r}{a_{v\tau}} 
    \frac{e^{\frac{r-R_{v\tau}}{a_{v\tau}}}}{\left( 1+ e^{\frac{r-R_{v\tau}}{a_{v\tau}}} \right)^2}.
\end{equation}

In the case of $^{212}$Po and $^{104}$Te  the two-particle  amplitudes of Eq.~(\ref{twop}) were obtained exactly in the  Tamm-Dancoff approximation \cite{1964Lane,2004Ring}:
\begin{equation} \label{eq.x}
 X_{\nu}^{\tau}=N_0 \frac{f(\nu,\tau)}{2 \epsilon^\tau_\nu - E^\tau_0},
\end{equation}
where $\epsilon^\tau_\nu$  are  s.p. energies,  $E^\tau_0$ is the correlated two-particle energy and $N_0$ is the normalization constant fixed by the condition $\sum_{\nu} \left( X_{\nu}^{\tau} \right)^2=1$.

The parameters $R_{v\tau}$ and $a_{v\tau}$ defining  the radial form factor (\ref{ff}) for $^{210}$Pb and $^{210}$Po were chosen to reproduce  the wave functions used by Harada \cite{1961Harada}. Since such data are not available for $^{102}$Sn and $^{102}$Te, in this case we adopted  the values of the WS potential for $^{100}$Sn shown in  Table~\ref{table.ws}.   
The pairing strength $G_{\tau}$ was adjusted to fit the experimental two-nucleon separation energies $S_{2\tau}$ through the dispersion relation
\begin{equation}
  \frac{1}{G_\tau} = \sum_\nu \frac{ f^2(\nu,\tau)}{2 \epsilon_\nu^\tau - E^\tau_0}.
\end{equation}
Since the proton-unbound nucleus  $^{102}$Te is not known experimentally, for this system we adopted the value of $S_{2p}=-2.14$\,MeV obtained by extrapolating down from the heavier Te isotopes  \cite{nndc}. This value is in  reasonable agreement with recent phenomenological estimates \cite{2002Isakov}. 
Table~\ref{table.2sp} lists the parameters of the residual interaction used in our study.
\begin{table}[htb]
\caption{\label{table.2sp} Parameters $R_{v\tau}$ and $a_{v\tau}$ of the residual interaction (\ref{vres}). The last column lists the value of $S_{2\tau}$ that has been used to constrain the pairing strength $G_{\tau}$ for various configuration spaces considered.}
 \begin{ruledtabular}
\begin{tabular}{cccc}
  nucleus           & $R_v$ (fm) & $a_v$ (fm) & $S_{2\tau}$ (MeV) \\
\hline
 $^{210}$Pb & 7.525 & 0.70 & 9.123  \\
 $^{210}$Po & 5.451 & 0.75 & 8.783 \\[2pt]
 $^{102}$Sn & 5.895 & 0.70 & 24.3 \\
 $^{102}$Te  & 5.895 & 0.70 &  $-$2.14 \\
\end{tabular}
 \end{ruledtabular}
\end{table}

\subsection{Configuration space}
To study the dependence of the  formation amplitude on the size of valence space, and to compare with previous work, we considered  several model spaces. Those used in the description of  the alpha decay of $^{212}$Po are given in 
Table \ref{table.msPb}. 
The model space M0 contains only one valence shell. The space M1 contains one  major shell, including  the unusual-parity intruder orbit. The model space M2 is that  used by Harada \cite{1961Harada}. The model space M3 is that of Glendenning and Harada \cite{1965Glendenning}. Finally, M4 is  the extended shell model space employed by Tonozuka and Arima.
\begin{table}[ht]
\caption{\label{table.msPb} Model spaces used in this work to describe  $^{212}$Po alpha decay.}
\begin{ruledtabular}
\begin{tabular}{|c|c|c|}
   Model       & Neutron States & Proton States \\
\hline
   M0   &  $1g_{9/2}$     &  $0h_{9/2}$ \\
\hline
   M1        &  $1g_{9/2},0i_{11/2},2d_{5/2},0j_{15/2}$     &  $0h_{9/2},1f_{7/2},0i_{13/2},2p_{3/2}$ \\
                          &  $3s_{1/2},2d_{3/2},1g_{7/2}$                     &  $1f_{5/2},2p_{1/2}$ \\ 
\hline
   M2         & $1g_{9/2},0i_{11/2},2d_{5/2}$                     &  $0h_{9/2},1f_{7/2},0i_{13/2}$ \\
\hline
   M3        & $1g_{9/2},0i_{11/2},2d_{5/2},0j_{15/2}$     &  $0h_{9/2},1f_{7/2},0i_{13/2}$ \\
\hline
  M4       & $1g_{9/2},0i_{11/2},2d_{5/2},0j_{15/2}$     & $0h_{9/2},1f_{7/2},0i_{13/2},2p_{3/2}$ \\
           & $3s_{1/2},2d_{3/2},1g_{7/2},1h_{11/2}$      &  $1f_{5/2},2p_{1/2},1g_{9/2},0i_{11/2}$ \\ 
           & $0j_{13/2}$                                 &  $0j_{15/2},2d_{5/2},3s_{1/2},1g_{7/2}$ \\
           &                                             &  $2d_{3/2},1h_{11/2},0j_{13/2},1h_{9/2}$ \\ 
\end{tabular}
\end{ruledtabular}
\end{table}
The model spaces used to describe $^{104}$Te alpha decay are shown in Table~\ref{table.msSn}; M1 consists of  one  major shell, including  the unusual-parity intruder orbit, while M4 consists of states with width less than $1$\;MeV.
\begin{table}[ht]
\caption{\label{table.msSn} Model spaces used in this work to describe $^{104}$Te alpha decay.}
\begin{ruledtabular}
\begin{tabular}{|c|c|c|}
  Model              & Neutron States & Proton States \\
\hline
   M1                   & $0g_{7/2},1d_{5/2},2s_{1/2},1d_{3/2}$     &  $0g_{7/2},1d_{5/2},2s_{1/2},1d_{3/2}$ \\
                      &  $0h_{11/2}$                                             &  $0h_{11/2}$ \\ 
\hline
   M4                   & $0g_{7/2},1d_{5/2},2s_{1/2},1d_{3/2}$     &  $0g_{7/2},1d_{5/2},2s_{1/2},1d_{3/2}$ \\
                      &   $0h_{11/2},1f_{7/2},2p_{3/2},2p_{1/2}$    &  $0h_{11/2},1f_{7/2},0h_{9/2},0i_{13/2}$ \\
                      &   $0h_{9/2},1f_{5/2},0i_{13/2}$             & \\ 
\end{tabular}
\end{ruledtabular}
\end{table}

\subsection{Wave functions} \label{sec.bcswf}
For the alpha formation amplitude in  $^{212}$Po  discussed in Sec.~\ref{sec.applications.ho} we considered the model spaces M2, M3 and M4. The wave function amplitudes in  M2 were taken from Refs.~\cite{1960Banerjee,1961Harada}. 
For calculations in M3, we took the $T$=1 seniority-zero amplitudes  of Ref.~\cite{1965Glendenning} and renormalized them accordingly.
For calculations in the extended space M4, we used the renormalized amplitudes of Ref.~\cite{1979Tonozuka}; here we retained only  configurations having  width smaller than 1\,MeV. 
The comparison between $^{212}$Po and $^{104}$Te discussed in Sec.~\ref{sec.applications.compare} was carried out in the model spaces M1 and M4. The corresponding  wave functions were calculated in the two-particle approximation described in Sec.~\ref{subsec.force}, except for $^{212}$Po in the M4 model space, where Ref.~\cite{1965Glendenning} was used instead.

\subsection{Penetration factor}\label{sec.gammasp}

The s.p. alpha width $\Gamma_0^{sp}$ has been  obtained from the current expression (\ref{eq.gammasp}). The alpha-core potential was assumed to be of a WS+Coulomb form with the parameters of Ref.~\cite{1976DeVries}: $r_0=R_c=1.315$\,fm, $a=0.65$\,fm. 
The strength of the WS potential has been adjusted to reproduce the
measured $Q_\alpha$ value corrected by the electron screening term \cite{1954Thomas,Toth1960,1964Mang,1974Rasmussen,Hatsukawa1990}:
\begin{equation}
Q_\alpha = E_\alpha \frac{A_P}{A_D} + \Delta E_{\rm sc}
\end{equation}
where 
\begin{equation}\label{escr}
\Delta E_{\rm sc} = 65.3 Z_P^{1.4}-80 Z_P^{0.4}~(\rm eV).
\end{equation}

For $^{212}$Po, $E_\alpha = 8.785$\,MeV \cite{nndc} and $\Delta E_{\rm sc}=31.8$\,keV; hence, $Q_\alpha= 8.986$\,MeV. The g.s. alpha decay of  $^{104}$Te has not been observed. For that reason, we took the value  $Q^{\rm BE}_\alpha$=5.135\,MeV extrapolated down from the binding energy differences in   $^{108}$Te (3.445\,MeV) and $^{106}$Te (4.290\,MeV) \cite{RykGrz11}. By adding the screening correction $\Delta E_{\rm sc}=16.1$\,keV, we arrived at 
$Q_\alpha = 5.151$\,MeV.
The resulting WS 
potential strength is  $V_0=143.49$\,MeV for $^{212}$Po and $149.64$\,MeV for
 $^{104}$Te.
 
The Gamow wave functions were obtained by means of the code  {\sc anti} \cite{1995Ixaru}. The   complex energy of the metastable alpha state is $\mathcal{E}_\alpha=(8.986-i0.632 \times 10^{-13})$\,MeV for $^{212}$Po and
$\mathcal{E}_\alpha=(5.151-i0.814 \times 10^{-13})$\,MeV for $^{104}$Te. The outgoing spherical Coulomb-Hankel function $H^+$ was calculated using the code \cite{1985Thompson}.

\subsection{Calculation of the spectroscopic factor}

The radial integration in the expressions for the spectroscopic factor (\ref{eq.s2}) and the formation amplitude in the normalized SGB (\ref{eq.gnu}) have been carried out  using $200$ Gauss-Legendre mesh-points with the maximum radius of  $20$ fm. 

\begin{figure}[htb]
 \includegraphics[width=0.45\textwidth]{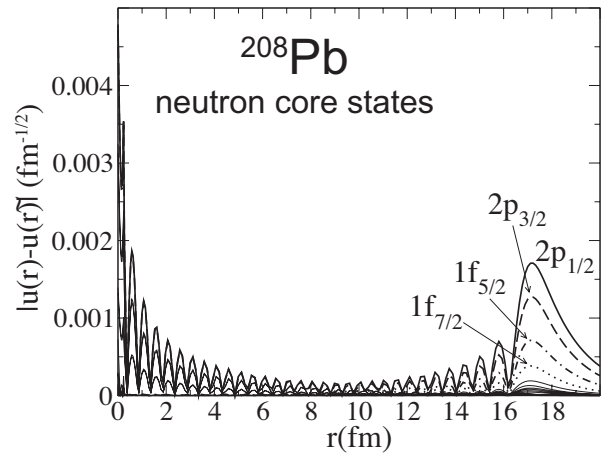}
 \caption{ \label{fig.udiff} $u_{{\rm diff}}(r)=|u(r)-\tilde{u}(r)|$ for  the  neutron core states in $^{208}$Pb for $\Delta R=0.5$\,fm, $M=30$, and  $R_{\textrm{max}}=15$\,fm.}
\end{figure}

The s.p. core wave functions  entering Eq. (\ref{eq.nkkp2}) are those of the  s.p. Hamiltonian of Table~\ref{table.ws}.
The radial mesh $R_k$ defining the normalized SGB (\ref{eq.tildeF}) was taken at equidistant points $R_k=k\, \Delta R$. In order to determine the step $\Delta R$ we expanded the s.p. core states $u(r)$ in the normalized SGB: $\tilde{u}(r) \equiv \sum_{k=1}^M \; a_k\; [r \tilde{F}_0(r,R_k)]$. Under the condition that $u_{{\rm diff}}(r)=|u(r)-\tilde{u}(r)|| < 0.005$\,fm$^{-1/2}$ we found that $0.44\,\rm{fm} \lesssim \Delta R \lesssim 0.57$\,fm  and $R_{\textrm{max}} \gtrsim 14$\,fm. For this range of $\Delta R$  and $R_{\textrm{max}}$ the normalized SGB is orthonormal with an accuracy better than $10^{-9}$. To illustrate the quality of the resulting  expansion, Fig. \ref{fig.udiff} shows $u_{{\rm diff}}(r)$ for the neutron core states in $^{208}$Pb.

\begin{figure}[htb]
 \includegraphics[width=0.45\textwidth]{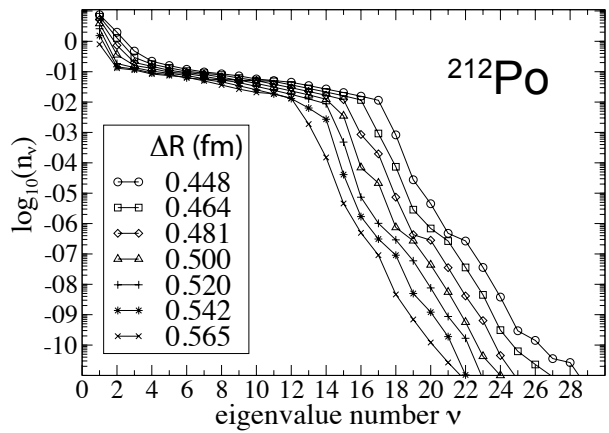}
 \caption{ \label{fig.nPo} Eigenvalues of the norm kernel (\ref{eq.nk}) for $^{212}$Po for $R_{\textrm{max}}=13$ fm for different values of $\Delta R$.}
\end{figure}

\begin{figure}[htb]
 \includegraphics[width=0.45\textwidth]{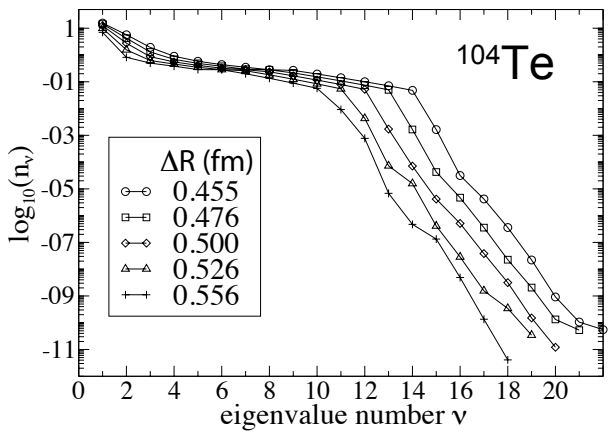}
 \caption{ \label{fig.nTe} Similar as in Fig.~\ref{fig.nPo} except for  $^{104}$Te and  $R_{\textrm{max}}=10$ fm.}
\end{figure}

To calculate the modified formation amplitude $G(R)$, one needs to determine the  eigenvalue cutoff $n_{\rm min}$. To this end, we show in Figs.~\ref{fig.nPo} and \ref{fig.nTe} typical distribution of the eigenvalues $n_\nu$ of the norm kernel (\ref{eq.nk}) for  $^{212}$Po  and $^{104}$Te, respectively,  for different values of $\Delta R$. One may observe that a significant fraction of them accumulate at zero \cite{1990Bonche,2004Ring}. To eliminate these spurious eigenvectors, we  define the cutoff at the value where the eigenvalue distribution changes slope. For $^{212}$Po and $^{104}$Te this happens at $n_\nu$ around $10^{-3}$. Consequently, in our calculations, we adopt the cutoff value of $n_{\rm min}=0.001$.

\begin{figure}[htb]
 \includegraphics[width=0.45\textwidth]{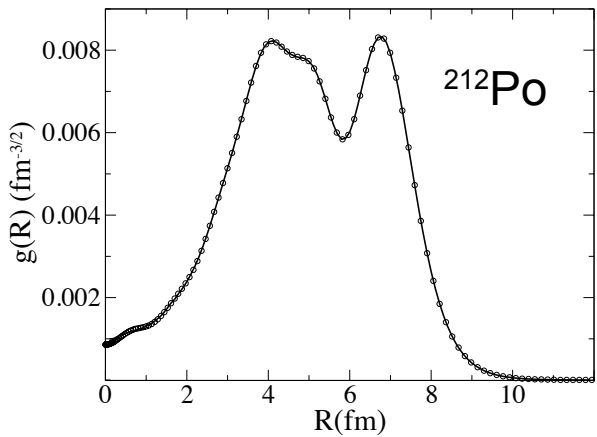}
 \caption{ \label{fig.gexp} Formation amplitude $g(R)$ for $^{212}$Po in the M4 model space expanded in eigenfunctions of the norm kernel for $\Delta R=0.500$ fm and  $R_{\textrm{max}}=13$ fm.}
\end{figure}

The eigenfunctions $u^L_\nu(R)$ of the norm kernel (\ref{eq.ur})  are orthonormal with an accuracy of  $10^{-10}$ for all eigenvalues. The eigenfunctions with  $n_\nu < n_{\rm min}$ oscillate inside the nuclear volume  and vanish outside the surface region.  To further check the quality  of  $u^L_\nu(R)$ we compute expression  (\ref{eq.gbig2}) by assuming  $n_{\rm min}=0$ and $n_\nu=1$ for all $\nu$. In this case, Eq.~(\ref{eq.gbig2}) formally reduces  to $g(R)$. Figure \ref{fig.gexp} shows
$g(R)$ for $^{212}$Po calculated in this way. The agreement with the original formation amplitude is excellent, except for a small deviation close to $R=0$ and a  small oscillation around and beyond the nuclear surface, which is not visible in the scale of  Fig.~\ref{fig.gexp}.

\begin{figure}[htb]
 \includegraphics[width=0.45\textwidth]{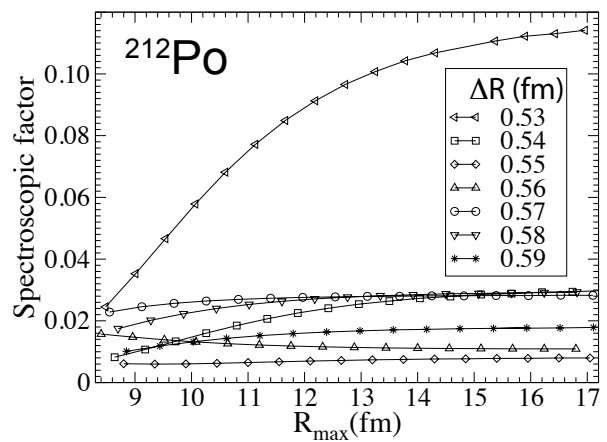}
 \caption{\label{fig.svsrmax} Convergence of $\mathcal{S}$ for $^{212}$Po (model space M4) as a function of $R_{\textrm{max}}$ of the normalized SGB for different values of  $\Delta R$ (with $n_{\rm min}=0.001$.)}
\end{figure}

Next we study the sensitivity of $\mathcal{S}$ to the choice of $R_{\textrm{max}}$, $\Delta R$, and $n_{\rm min}$. For this analysis we relax the condition for $u_{{\rm diff}}(r)$ in order to access a wider range of $\Delta R$. First, we study the sensitivity of $\mathcal{S}$ as a function of $R_{\textrm{max}}$ for various values of  $\Delta R$. Figure \ref{fig.svsrmax} shows the result for $0.53 \rm{fm} \leq \Delta R \leq 0.59$ fm for $^{212}$Po in the model space M4 and $n_{\rm min}=0.001$. Except for a small value of $\Delta R=0.53$\,fm, which does not produce stable results, a plateau in  $R_{\textrm{max}}$ is reached around 14\,fm. 

\begin{figure}[htb]
 \includegraphics[width=0.45\textwidth]{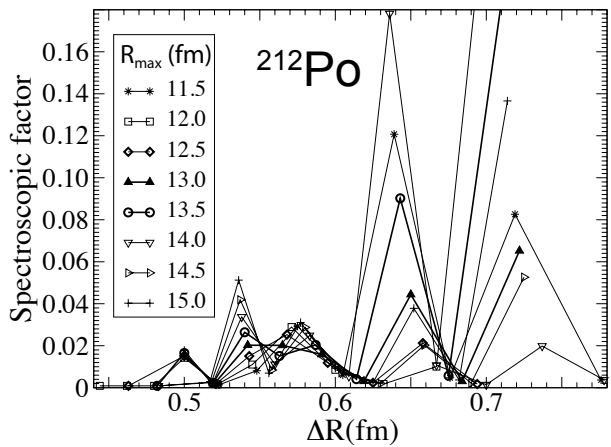}
 \caption{\label{fig.svsDR} Similar as in Fig.~\ref{fig.svsrmax} but as a function of the step size $\Delta R$  for different values of $R_{\textrm{max}}$.}
\end{figure}

The dependence of $\mathcal{S}$ on $\Delta R$ displayed in Fig.~\ref{fig.svsrmax} reflects the fact that for too small values of the step
the basis functions become numerically linearly dependent, while for too large $\Delta R$'s the basis cannot capture  high Fourier components \cite{1976Fliessbach1,1990Bonche,2004Ring}.
Figure~\ref{fig.svsDR} shows $\mathcal{S}$ for $^{212}$Po in the model space M4 and $n_{\rm min}=0.001$ as a function of $\Delta R$. In general, appreciable oscillations of $\mathcal{S}$ can be seen except for the ``safe" region $0.54\,\rm{fm} \leq \Delta R \leq 0.59$ fm, where results weakly depend on $R_{\textrm{max}}$.

\begin{figure}[htb]
 \includegraphics[width=0.45\textwidth]{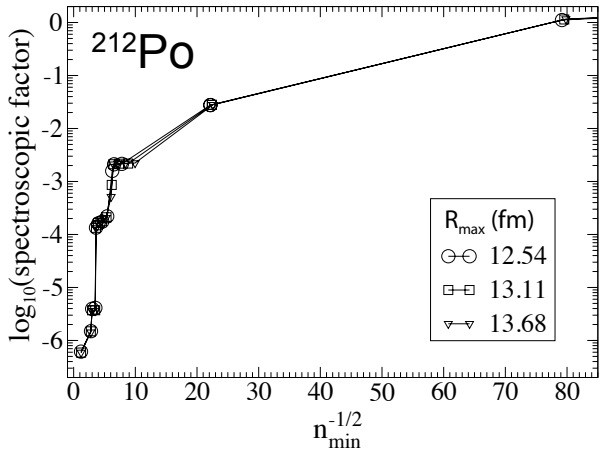}
 \caption{\label{fig.svsnmin} Similar as in Fig.~\ref{fig.svsrmax} but as a function of  $n_{\rm min}$ for  different values of $R_{\textrm{max}}$ and $\Delta R=0.57$.}
\end{figure}

Finally, Fig.~\ref{fig.svsnmin} shows the behavior of $\mathcal{S}$ as a function of the eigenvalue cutoff $n_{\rm min}$ for $\Delta R=0.57$\,fm. The cutoff used in Figs. \ref{fig.svsrmax} and \ref{fig.svsDR} corresponds to $n^{-1/2}_{\rm min}=(0.001)^{-1/2} \approx 31.5$.

\subsection{Integral over intrinsic coordinates}
The multidimensional integral (\ref{eq.inp}) depends on the nucleonic coordinates, which are parametrized in terms of intrinsic  variables through Eqs.~(\ref{eq.rhoandr}) and (\ref{eq.cos}). The integration over $\tilde{\varphi}_i$ can easily be done analytically. Since the coordinates of particles 1 and 2  depend only on the relative coordinates 1 and 3, and the particle coordinates 3 and 4  depend only on the relative coordinates 2 and 3,  one can greatly simplify the remaining six-dimensional integral by making first the integration over the relative coordinates 1 and 2 and then the integration over the coordinate 3: 
\begin{equation}
    \int d\gras{\rho}_3 \left[ \dots \left(\int \dots d\gras{\rho}_1\right)  \left( \int \dots d\gras{\rho}_2 \right) \right].
\end{equation}
The integration has been  carried out using the Gauss-Legendre quadrature
using 10 points for the radial integrals and 8 points for the 
the angular coordinates. This guarantees the convergence up to the fourth significant digit.

\section{Reduced width for $^{212}$P\lowercase{o}} \label{sec.applications.ho}

\subsection{Single-$j$ configuration}

Following Rasmussen \cite{1963Rasmussen}, it is instructive to compute 
relative reduced widths 
assuming a pure single-$j_n$ shell model orbital assignment for the neutron
pair, while  the proton pair fills the  $0h_{9/2}$ shell.
For
simplicity,  the results are expressed relative to the $^{210}$Po reference (a neutron pair in  $2p_{1/2}$).

In the delta-function approximation of Sec.~\ref{subsec.formalism.delta}, the ratio $r_\delta$ of the reduced widths  is given by a simple expression \cite{1963Rasmussen}:
\begin{equation} \label{eq.ratiodelta}
  r_{\delta} = \frac{\gamma^2_{j_n}}{\gamma^2_{2p_{1/2}}}=\frac{2j_n+1}{2} \; \left( \frac{u_{j_n}(R)}{u_{2p_{1/2}}(R)} \right)^4.
\end{equation}
In a more general case expressed by Eq.~(\ref{eq.f0}), the ratio $r$ depends on the proton wave function: 
\begin{equation}  \label{eq.ratio}
  r = \frac{\gamma^2_{j_n,0h_{9/2}}}{\gamma^2_{2p_{1/2},0h_{9/2}}} =  \frac{2j_n+1}{2} \; \left( \frac{I_{j_n,0h_{9/2}}(R)}{I_{2p_{1/2},0h_{9/2}}(R)} \right)^2.
\end{equation}
Table~\ref{table.ratio} compares the ratio $r_\delta$ given by Eq.~(\ref{eq.ratiodelta}) using the WS wave functions with that of Table~I of Rasmussen  \cite{1963Rasmussen} based on the rounded square well potential of Blomqvist and Wahlborn \cite{1960Blomqvist} for several neutron configurations at $R=9.5$\,fm. We find excellent agreement between these two calculations, and we checked that  
this agreement also holds for $R=9.0$\,fm. This is not surprising as both calculations employ finite-depth potentials.
\begin{table}
\caption{ \label{table.ratio} Single-$j$ alpha reduced width ratios  at $R=9.5$\, fm. Shown are: $r_\delta$ of Ref.~\cite{1963Rasmussen}, $r_\delta$ of Eq.~(\ref{eq.ratiodelta}), $r$ of Eq.~(\ref{eq.ratio}), and $r$ of Ref.~\cite{1962Zeh}.}
\begin{ruledtabular}
\begin{tabular}{ccccc}
 Orbital $j_n$    &  $r_\delta$ \cite{1963Rasmussen} & $r_\delta$  & $r$  &  $r$ \cite{1962Zeh}  \\
\hline
 $0i_{13/2}$ & 0.44							  &  0.46    &    0.20 & 0.10\\
 $0i_{11/2}$ & 0.32							  &  0.34  &    0.21  &  0.08\\
 $1g_{9/2}$  & 7.50							 &  7.50     &  6.50  & 3.73     \\
 $1f_{5/2}$   & 0.73							 &  0.74  &  0.58  &  0.55   \\
 $2p_{3/2}$  & 1.89							 &  1.89     &   1.73  &  1.89    
\end{tabular}
\end{ruledtabular}
\end{table}
The fourth column of Table \ref{table.ratio} displays  the ratio $r$ given by 
Eq.~(\ref{eq.ratio}) using the WS wave functions; they are compared with the h.o. values of Ref.~\cite{1962Zeh} (last column). It is seen that h.o. calculations
underestimate  WS values for high-$j$ orbits by 
a factor two-to-three.

It has been early  recognized \cite{1960Mang,1963Rasmussen} that the delta-function  approximation  overestimates the contributions of high-$j$ orbitals. One can see it clearly by comparing the values of $r_\delta$  of Eq.~(\ref{eq.ratiodelta}) with those of $r$ (\ref{eq.ratio}), i.e., the third and fourth columns of Table \ref{table.ratio}. To cure this deficiency,
a correction factor $B_\nu$  (\ref{Bcorr})  was introduced \cite{1963Rasmussen} in Eq.~(\ref{Bcorr2}) that depends on the relative angular momentum.

\subsection{Enhancement due to configuration mixing}\label{enhan}

As   was first shown by Harada \cite{1961Harada}, the reduced width  at the surface region is strongly enhanced by the configuration mixing  because 
contributions from various shell model orbits add  coherently.
To assess the effect of collective enhancement due to the configuration mixing,
we carried out calculations in the M2 space. For $R=8$\,fm, 
our WS calculations yield the enhancement factor of $\zeta=8.5$ with respect to the valence-shell configuration M0. This is to be compared with $\zeta=11$ obtained 
in the delta-function  approximation;  $\zeta=10$ obtained by Rasmussen \cite{1963Rasmussen}; and $\zeta=5.5$ of  Harada \cite{1961Harada} using h.o. wave functions. 

For the model space M3 of Glendenning and Harada \cite{1965Glendenning},
obtained by adding the intruder neutron state $0j_{15/2}$ to M2, we obtain $\zeta=21$. This should be  compared with $\zeta=24$ obtained 
in the delta-function  approximation and  $\zeta=30$ obtained in Ref.~\cite{1965Glendenning} (also within the delta-function  approximation) using a fairly rich wave function that also includes proton-neutron correlations and $J>0$ two-particle couplings. It is worth noting  that our enhancement is around $80\%$ of that by Glendenning and Harada, and that  the seniority-zero component   in  their wave function is also 80\%.

\subsection{Extended shell model space}

Due to the strong collective enhancement of the reduced width due to configuration mixing, it is important to consider extended shell-model space by taking into account higher-lying  orbitals~\cite{1979Tonozuka}. For finite-depth 
shell-model potentials, such as the WS potential used in this study,  this necessitates a proper treatment of the particle continuum. An appropriate representation to deal with the continuum space is the complex Berggren ensemble representing bound and unbound s.p. states \cite{1968Berggren,Michelrev}.

Here we consider the large configuration space M4 of Tonozuka and Arima \cite{1979Tonozuka}, i.e., 
all s.p.  orbits up to $N = 7$ harmonic oscillator shell except for broad resonances  with widths greater than 1 MeV. The shell-model amplitudes were taken from  Ref.~\cite{1979Tonozuka} and renormalized to the reduced model space. For the sake of comparison with Ref.~\cite{1979Tonozuka}, we consider
the relative reduced width
\begin{equation}\label{thetaw}
\theta^2(R)=\frac{\gamma^2(R)}{\gamma^2_W(R)},
\end{equation}
where  $\gamma^2_W(R)=\frac{3 \hbar^2}{2 \mu R^2}$ is the  Wigner limit  \cite{1958Lane}.

Table \ref{table.tonozuka} compares our WS results for $\theta^2(R)$ with those   of Ref.~\cite{1979Tonozuka} obtained in the h.o. basis for several values of $R$.
\begin{table}[htb] 
\caption{\label{table.tonozuka} The relative reduced width $\theta^2$  (\ref{thetaw})
obtained in Ref.~\cite{1979Tonozuka} and this work.}
\begin{ruledtabular}
\begin{tabular}{cccc}
 Model space &  $R$ (fm)  & Ref.~\cite{1979Tonozuka}  & This work \\
 \hline
 M0          &  8.4        &    $6.3 \times 10^{-6}$    &    $0.60 \times 10^{-6}$     \\
 M3          &  8.5        &    $4.4 \times 10^{-5}$    &    $0.48 \times 10^{-5}$     \\
 M4          &  9.0        &    $2.9 \times 10^{-4}$    &    $0.41 \times 10^{-4}$    
\end{tabular}
\end{ruledtabular}
\end{table}
Generally,  the reduced width obtained in the WS model is about one order of magnitude smaller than that in the h.o. basis. This is because the h.o. basis knows nothing about the particle thresholds, and the radial behavior at large distances is solely determined by the oscillator length. For that reason, calculations based on the h.o. wave functions show large sensitivity to this parameter \cite{1962Harada}.

\begin{figure}[htb]
 \includegraphics[width=0.45\textwidth]{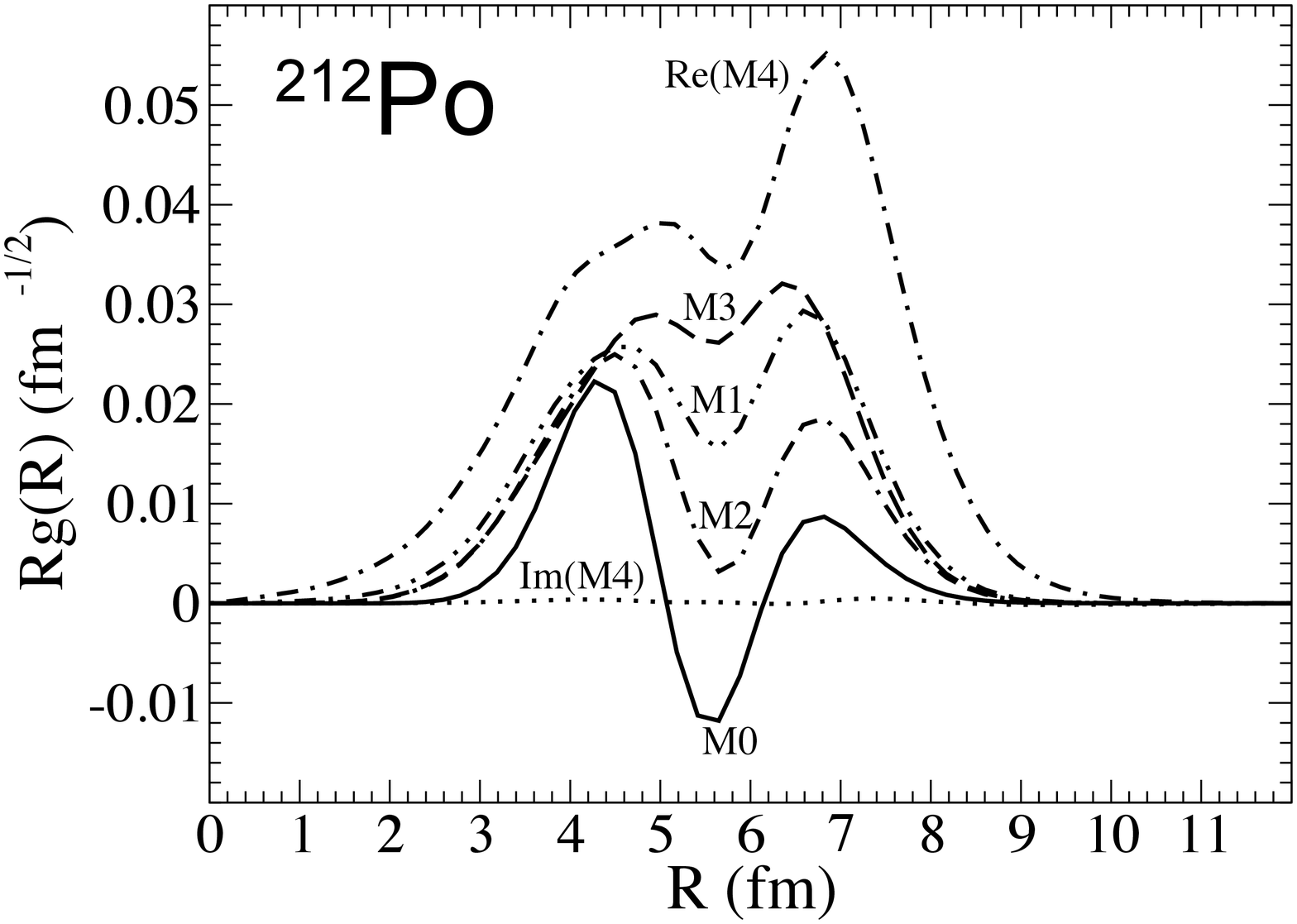}
 \caption{ \label{fig.fl} Formation amplitude $g(R)$ for $^{212}$Po obtained in this work in the model spaces M0 to M4 as defined in table \ref{table.msPb}. The imaginary part of the formation amplitude in M4 (dotted line) is also shown.}
\end{figure}
The formation amplitude obtained in this work is shown in 
Fig. \ref{fig.fl} for the configuration spaces M0, M1, M2, M3, and M4. 
Compared with the formation amplitudes of Ref.~\cite{1979Tonozuka},  the maximum 
of the formation amplitudes obtained in 
the WS model are significantly larger, and appear at lower values of $R$,  than in the h.o. model. Also
the overall shape of the formation amplitude is very different in the two cases. A  characteristic two-humped shape of $g(R)$ calculated in M4 resembles the formation amplitude $G(R)$ obtained in Refs.~\cite{1976Fliessbach,1979Tonozuka}. 
A similar result was also obtained in Refs.~\cite{1989Okabe,1992Varga}.
It is indeed interesting to see that a two-humped behavior of  the formation amplitude for $^{212}$Po has been obtained by considering large configuration space and the Berggren ensemble of the WS potential.

Figure~\ref{fig.fl} also shows that the formation amplitude  in the M4 model space has a small imaginary part. This is because our calculations are carried out in the pole approximation that ignores the non-resonant continuum \cite{1993Beggren,2003IdBetan,2003Michel,Michelrev}. This  spurious component of $g(R)$ results in a very small imaginary contribution to the reduced width, which can be safely neglected considering the expected accuracy of our model.

\section{Absolute alpha-decay width of $^{212}$P\lowercase{o}} \label{sec.aw}

The  g.s. alpha-decay  width of  $^{212}$Po has been determined in the  seniority-zero approximation using three different models spaces listed in  Table \ref{table.msPb}: M0, M3, and M4. The corresponding four-particle shell-model wave function contains one configuration in the M0 space, 12 configurations in M3, and 144 seniority zero configurations in M4.

\begin{figure}[htb] 
 \includegraphics[width=0.45\textwidth]{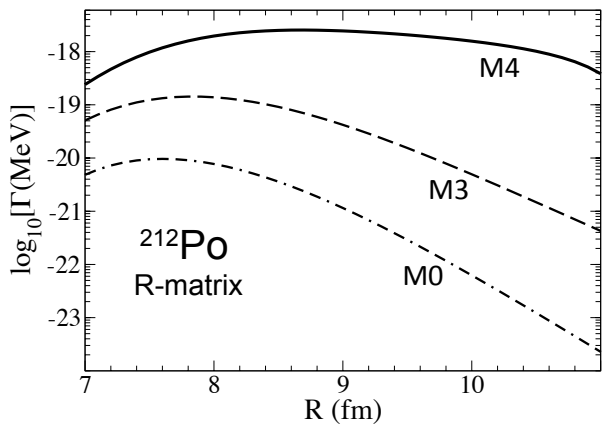}
 \caption{\label{fig.gamma} Dependence of the absolute  alpha-decay width 
 (\ref{Grmatrix}) of  $^{212}$Po  on the R-matrix channel radius $R$
for  three different model spaces M0, M3 and M4.}
\end{figure}

The absolute width from Eq.~(\ref{Grmatrix}) should not depend on the channel radius  $R$. However, in  R-matrix studies involving approximations, such as the  one-channel R-matrix treatment, this  condition cannot be met 
\cite{2010Descouvemont}. Therefore, in practical calculations, in which
the dependence of $\Gamma_L$ on $R$ around the nuclear surface is small relative to the appreciable $R$-dependence of the formation amplitude, one is trying to meet the plateau condition for $\Gamma_L(R)$ in which the absolute width varies weakly  around the nuclear surface \cite{1988Insolia}.  Figure~\ref{fig.gamma} shows  the dependence of the R-matrix width (\ref{Grmatrix}) on the channel radius.
It is seen that the plateau condition is met only in the case of the extended configuration space M4 involving particle continuum. Here, we find  a fairly weak variation of $\Gamma(R)$  between 7\,fm and 11\,fm. 
 
As seen in Fig.~\ref{fig.gamma}, and discussed in  Sec.~\ref{enhan} and Refs.~\cite{1962Harada,1979Tonozuka,2010Qi}, the width strongly increases with the size of the shell-model space. Indeed, in the surface region, 
$\Gamma(R)$ obtained in M3 shows an enhancement $\sim$15 with respect to M0, and in the extended space M4 the enhancement is $\sim$260. Compared to experimental value, however, the width obtained in M4 is still 600 times smaller than the experimental value  $\Gamma_{\rm exp}=0.153 \times 10^{-14}$\, MeV \cite{nndc}.

\begin{figure}[htb]
\includegraphics[width=0.45\textwidth]{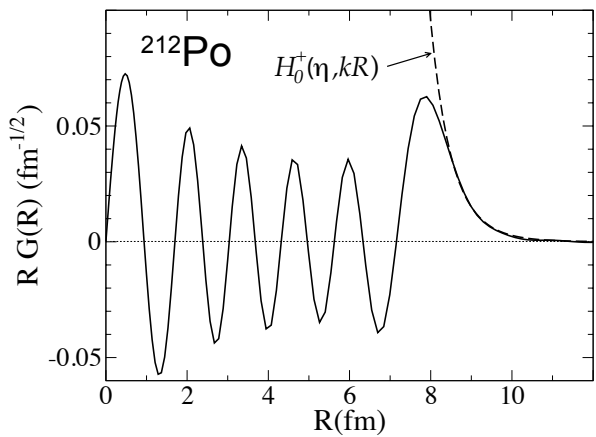}
 \caption{ \label{fig.gbig} Modified formation amplitude $G(R)$ of Eq.~(\ref{eq.gbig}) in the extended  model space M4 with $n_{{\rm min}}=0.001$, $\Delta R=0.56$ fm and $R_{\textrm{max}}=11.76$. Unlike $g(R)$, $G(R)$ properly accounts for the normalization and antisymmetrization of the decay channel. The asymptotic behavior of $G(R)$ is given by  the Coulomb-Hankel function at the alpha-decay energy $Q_\alpha=8.986$ MeV (dashed line).}
\end{figure}

A further enhancement in the reduced width is due to the antisymmetrization and normalization of the channel decay \cite{1975Fliessbach,1976Fliessbach}. This is achieved by replacing the standard formation amplitude $g(R)$ with the modified formation amplitude $G(R)$ of Eq.~(\ref{eq.gbig}). Figure~\ref{fig.gbig} shows $G(R)$ calculated in the M4 model space with $\Delta R=0.56$ fm, $R_{\textrm{max}}=11.76$ fm ($M=21$) and $n_{\rm min}=0.001$. A small oscillation at the tail of $G(R)$  can be seen. The amplitude of this oscillation, around the asymptotic behavior given by  $H_0^+(\eta,kR)$, varies very little with  $R_{\textrm{max}}$ for this value of $\Delta R$.  As discussed in, e.g., \cite{1976Fliessbach,1992VargaLovas,1998Lovas}, the behavior  of $g(R)$ and $G(R)$ is generally very different. This can be seen by comparing Figs.~\ref{fig.fl} and \ref{fig.gbig}. 

\begin{figure}[htb]
 \includegraphics[width=0.45\textwidth]{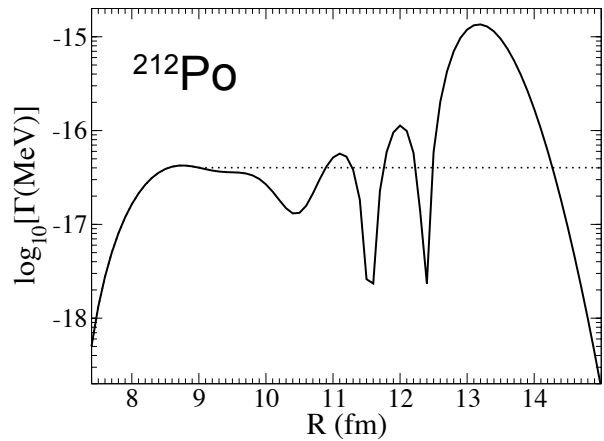}
 \caption{ \label{fig.gamma2} Absolute width from R-matrix expression (\ref{Grmatrix}) calculated in the M4 model space using the modified formation amplitude $G(R)$ of Fig.~\ref{fig.gbig}. At $R>9$\,fm, the result obtained by assuming $G(R) \propto H_0^+(\eta,kR)$  is marked by a dotted line.}
\end{figure}

The  absolute alpha-decay width obtained by using the R-matrix expression (\ref{Grmatrix}) with the formation amplitude $G(R)$ of Fig.~\ref{fig.gbig}  is shown in Fig.~\ref{fig.gamma2}. There appears a small plateau in the region of nuclear surface that corresponds to $\Gamma \approx 0.0042 \times 10^{-14}$ MeV. This value  is $\sim 36$ times smaller than $\Gamma_{\rm exp}$. 
At larger distances $R> 9$\,fm, the result is affected by spurious oscillations 
of $G(R)$ around $H_0^+(\eta,kR)$, i.e., it is  quite unreliable.

\begin{figure}
 \includegraphics[width=0.45\textwidth]{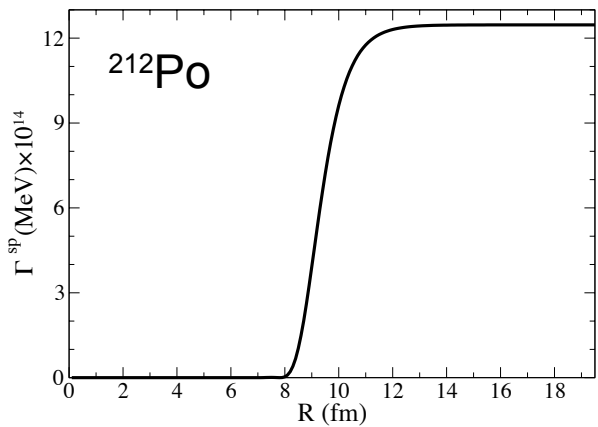}
 \caption{ \label{fig.gammasp} Single particle  width of $^{212}$Po from current expression (\ref{eq.gammasp}).}
\end{figure}

The absolute width  can also be obtained from  expression (\ref{Goverlap}), which involves the alpha-particle spectroscopic factor $\mathcal{S}$  and the s.p. decay width.  Figure~\ref{fig.gammasp} shows the result of  the current expression (\ref{eq.gammasp}) for $\Gamma^{sp}$ as a function of the channel radius. 
As discussed in Ref.~\cite{2004Kruppa},  $\Gamma^{sp}$ calculated this way
should be independent of $R$   if $R$  is
large enough. This is precisely what is seen in Fig.~\ref{fig.gammasp}: 
the s.p. width converges beyond the range of the WS potential to $\Gamma^{sp}=0.1247 \times 10^{-12}$ MeV, which is indeed very close to the value of $-2$Im$({\cal E}_\alpha)=0.1265  \times 10^{-12}$ MeV given by the imaginary part of the Gamow resonance.

Using the modified formation amplitude $G(R)$ of Fig.~\ref{fig.gamma2}, we  compute the spectroscopic factor $\mathcal{S} = 0.011$, which -- combined with the value of $\Gamma^{sp}$ above --  yields $\Gamma= 0.14 \times 10^{-14}$ \,MeV. Using $\Delta R=0.55$ fm we obtain $\mathcal{S} = 0.0080$ and  $\Gamma= 0.10 \times 10^{-14}$ \,MeV. Both these values are close to $\Gamma_{{\rm exp}}=0.153 \times 10^{-14}$\,MeV.

\section{Comparison between ground-state alpha decay of $^{212}$P\lowercase{o} and $^{104}$T\lowercase{e}} \label{sec.applications.compare}

To compare  absolute widths of $^{212}$Po and $^{104}$Te in a consistent way, we consider similar M1 and M4 model spaces for both nuclei.  The norm kernel eigenvalues $n_\nu$ do not depend on the model space in which $g(R)$ is calculated, so we  take the cutoff $n_{{\rm min}}=0.001$.

Let us begin with $^{212}$Po  by making a convergence analysis of $\mathcal{S}$ in the M1 model space as a function of $\Delta R$ and $R_{\rm{max}}$ (as in Fig. \ref{fig.svsrmax}). For $\Delta R=0.53,\, 0.54,\, 0.55$, and $0.56$ fm, we found $\mathcal{S} = 0.0041,\, 0.0011,\, 0.00030,\, 0.00032$, respectively. The resulting converged value $\mathcal{S} = 0.0003$ is  too small, as expected from Fig. \ref{fig.fl}. This deficiency is related to the poor quality of the interaction used to describe  $^{212}$Po in M1. To better understand this fact, let us take a look of the spectroscopic factor in terms of the spectral representation of the norm kernel,
\begin{equation} \label{eq.s3}
 \mathcal{S} =\sum_\nu \frac{g^2_\nu}{n_\nu},
\end{equation}
where the sum is truncated by the condition $n_\nu > n_{{\rm min}}$.
The summation range and eigenvalues $n_\nu$ are the same for  M1 and M4; the only difference comes from $g_\nu$. Because of the rapid oscillation of the eigenfunctions inside the nucleus, only the eigenfunctions which are peaked at and beyond the nuclear surface will contribute significantly to the sum. But -- because $g(R)$ in M1 is small in the surface region -- the overlap with those eigenfunctions is small, and this gives rise to a very reduced value of $\mathcal{S}$.

By making a similar analysis for $^{104}$Te in M1, we found $\mathcal{S} = 0.067,\, 0.024,\, 0.0066$, and $0.00046$ for $\Delta R=0.53,\, 0.54,\, 0.55$, and 0.56 fm,  respectively. In the model space M4 we found $\mathcal{S} = 0.21$, 0.088, 0.032, and 0.0051 for the same  values of $\Delta R$. 
Clearly, the convergence in $\mathcal{S}$ has not been achieved for $^{104}$Te. We would like to attribute this to the impact of the proton continuum on $g_\nu$, which results in increased oscillations of $G(R)$ in the surface area.
Table \ref{table.s} compares the  values of  $\mathcal{S}$ and the corresponding  absolute widths for $^{212}$Po and $^{104}$Te at $\Delta R=0.56$\,fm. (The single particle width for Te is $\Gamma^{sp}=0.162 \times 10^{-12}$ MeV.)
\begin{table}[htb]
 \caption{\label{table.s} Alpha decay spectroscopic factor and absolute width for $^{212}$Po and $^{104}$Te computed in the configuration spaces M1 and M4, with $n_{{\rm min}}=0.001$, $\Delta R=0.56$ and $R_{\textrm{max}}=11.76\,\rm{fm}\, (M=21)$.}
\begin{ruledtabular}
\begin{tabular}{c|cc|cc}
 Model Space  & 
 \multicolumn{2}{c|}{$\mathcal{S}$} &
  \multicolumn{2}{c}{$\Gamma\times 10^{14}$ MeV} \\
  & $^{212}$Po    & $^{104}$Te    & $^{212}$Po              & $^{104}$Te \\
\hline
 M1          & $0.00032$     & $0.00046$          & $0.0040$               & $0.0075$  \\
 M4          & $0.011$       & $0.0051$           & $0.14$                 & $0.083$
\end{tabular}
\end{ruledtabular}
\end{table}

\begin{figure}[htb] 
\includegraphics[width=0.45\textwidth]{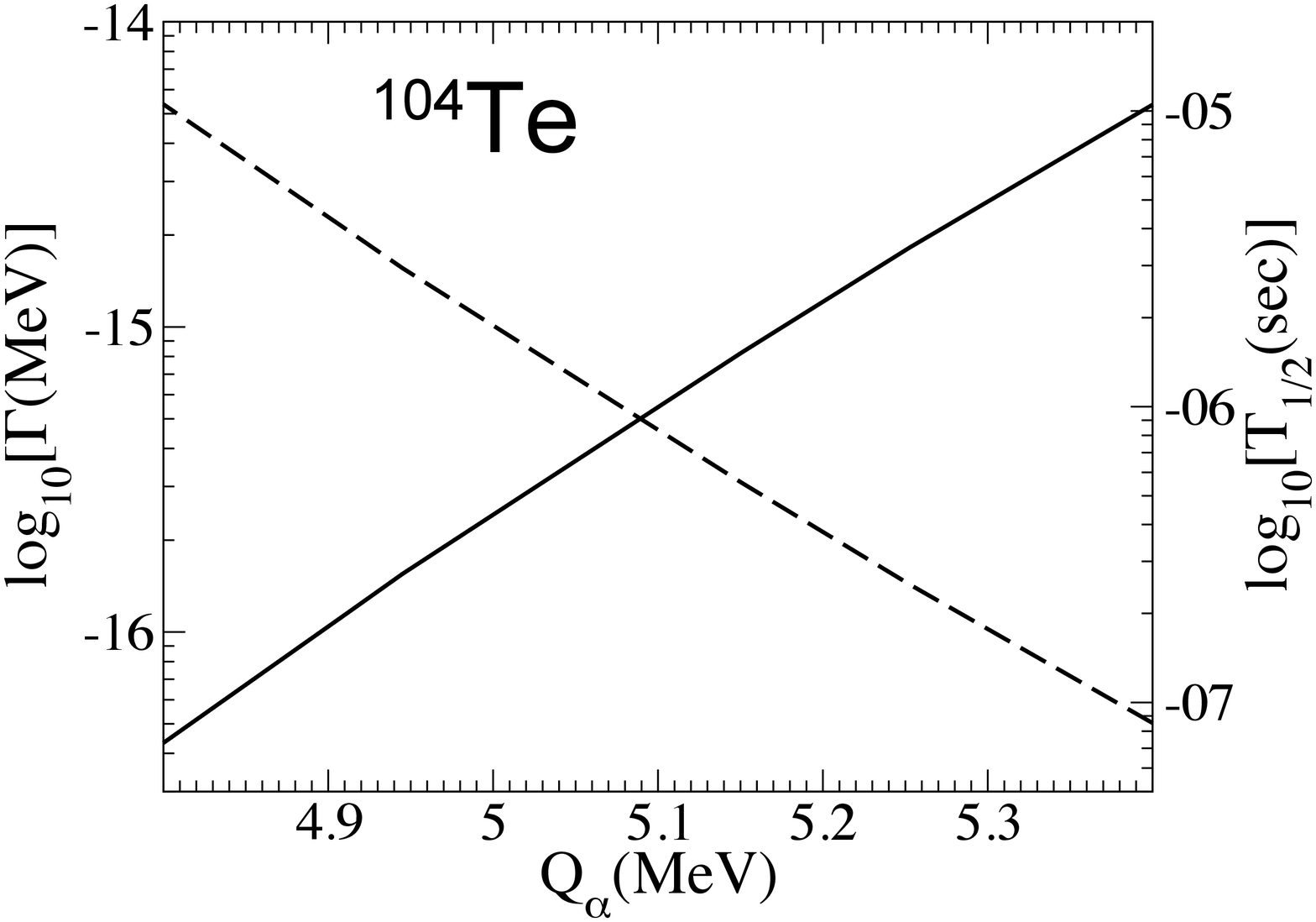}
 \caption{\label{fig.gammate} Ground-state alpha-decay width (left scale)  and half-life (right scale) in $^{104}$Te as functions of the decay energy.}
\end{figure}

It is interesting to compare our current results for $^{104}$Te with the estimates of phenomenological alpha-decay models based on semi-classical approximation \cite{Xu2006,Ren2006,Mohr07}. The assumed large value of $Q_\alpha=6.12$\,MeV in Ref.~\cite{Xu2006} results in a very short half-life of $7\times 10^{-11}$\,sec. The alpha-decay energies of 5.05\,MeV \cite{Ren2006} and 5.42$\pm$0.07\,MeV \cite{Mohr07} result in   $T_{1/2}\sim10^{-7}$\,sec and $\sim5\times 10^{-9}$\,sec, respectively, and these estimates are not inconsistent with our value (M4 model space) $T_{1/2}=5.5 \times 10^{-7}$\,sec ($Q_\alpha=5.151$\,MeV).
As the value of  $Q_\alpha$  in  $^{104}$Te is very uncertain, we show in Fig.~\ref{fig.gammate} the absolute width and half-life  $T_{1/2}$ as a function of $Q_\alpha$ for the model space M4.

Our predicted spectroscopic factors in M4 for $^{104}$Te  and $^{212}$Po are about 0.5\% and 1\%, respectively. As mentioned above,  a fairly small  value of $\mathcal{S}$ in $^{104}$Te could be a consequence of the proximity of the proton  continuum. Indeed, all the valence proton shells are resonances.
The small value of $\mathcal{S}$   in $^{104}$Te could also be attributed to the poor quality of the valence interaction assumed, and the neglect of the $T=0$ force. The effect of the proton-neutron interaction was examined in, e.g., Refs.~\cite{1961Harada,Sasaki1970} for $^{212}$Po and was found to be minor due to the fact that   neutrons and protons in $^{212}$Po occupy different shells. This is no longer true in the $N=Z$ nucleus $^{104}$Te, in which the major enhancement of $\mathcal{S}$ is expected due to $T=0$
correlations. Therefore, our predictions for $\mathcal{S}$ and $\Gamma$ in  $^{104}$Te  given in Table~\ref{table.s} should be considered as a very conservative lower limit.

\section{Conclusions} \label{sec.conclusions}

The g.s. alpha decay of $^{212}$Po has been studied within the complex-energy shell model framework with the Berggren ensemble of the average Woods-Saxon potential. We applied the pole approximation by considering   s.p. resonant states 
only. The overlap integral involving alpha-cluster nucleons was computed exactly, without resorting to the delta-function approximation. We considered the large valence space of Tonozuka and Arima that is necessary to produce the collective enhancement of the formation amplitude.

The absolute alpha-decay width was computed using the reduced width obtained in the framework of the R-matrix theory and also from the alpha spectroscopic factor. The latter approach yielded  results consistent with experimental value, but only after considering 
the antisymmetrization and normalization of the decay channel wave function.
The R-matrix estimate underestimates the experimental width by a factor of
$\sim 36$. 
The R-matrix expression depends on the asymptotic value of the formation amplitude that is very sensitive to the size of the configuration space. On the other hand, the reaction-theory expression (\ref{Goverlap}) involves the spectroscopic factor -- an integral quantity that depends less on the size of the basis used.
It is very encouraging to see that a reasonable agreement with the experimental width of $^{212}$Po has been obtained without  explicitly considering the alpha-cluster component in the wave function of the parent nucleus. In this context, we believe that the improved treatment of the particle continuum has been essential. 

We have  also provided an estimate of the alpha-decay rate in $^{104}$Te.
Unfortunately, due to the fact that the valence proton shells in this nucleus lie in the continuum, no fully convergent result has been achieved. We hope to improve the situation in the future by inclusion of the non-resonant continuum space that will remove some of the undesired oscillations in $G(R)$ at large distances.
In addition, since the residual interaction employed in our work neglects the proton-neutron components, and the wave function has a seniority-zero character based on $T=1$ 
nucleonic pairs, the predicted alpha width in this $N=Z$ nucleus should be viewed as a conservative low limit. Indeed, the inclusion of $T=0$ correlations is expected to increase the value of $\Gamma$ significantly.

The  calculations presented in this study should be considered as an important step   towards an improved  microscopic understanding of the alpha-decay process. Still, as this work demonstrates, further improvements are  needed.
The neglect of the non-resonant continuum, i.e., complex-energy scattering states in the Berggren ensemble, slightly violates the completeness relation at a one-body level. This results in small imaginary contributions to spectroscopic factors and reduced widths, and -- most importantly -- can  affect the behavior of formation amplitudes at very large distances. The second crucial development will be the use of large-scale shell model calculations, including realistic
$T=0$ and $T=1$ interactions, to compute wave function amplitudes. This will enable us to provide a more meaningful estimate of $^{104}$Te alpha decay rate.
The work in both directions is underway.

\begin{acknowledgments}
Useful discussions with Doru Delion, Torsten Fliessbach, Robert Grzywacz, Kiyoshi Kato, and Krzysztof Rykaczewski  are gratefully acknowledged. Special thanks to  Rezso Lovas for his insights on the norm kernel and his patience in explaining the underlying physics. This work was supported  by the Office of
Nuclear Physics,  U.S. Department of Energy under Contract No.
DE-FG02-96ER40963 and  by the National Council of Research PIP-77 (CONICET, Argentina).  
\end{acknowledgments}

%

\end{document}